# Modeling of Daily Precipitation Amounts Using the Mixed Gamma Weibull Distribution

**Hsien-Wei Chen**[a]

[a] Department of Civil Engineering and Applied Mechanics, McGill University, 817 Sherbrooke Street West, Montreal, Quebec H3A 0C3, Canada

**Abstract**

By recognizing that the main difficulty of the modeling of daily precipitation amounts is the selection of an appropriate probability distribution, this study aims to establish a model selection framework to identify the appropriate probability distribution for the modeling of daily precipitation amounts from the commonly used probability distributions, i.e. the exponential, Gamma, Weibull, and mixed exponential distributions. The mixed Gamma Weibull (MGW) distribution serves this purpose because all the commonly used probability distributions are special cases of the MGW distribution, and the MGW distribution integrates all the commonly used probability distributions into one framework. Finally, via the large sample inference of likelihood ratios, a model selection criterion can be established to identify the appropriate model for the modeling of daily precipitation amounts from the MGW distribution framework.

**Highlight**

- A model selection framework for daily precipitation amount modeling
- A probability distribution designed for hydrological purposes
- A solution to the model selection difficulty of daily precipitation amounts

## Introduction

The modeling of the precipitation time series for the daily scale is of greatest interest in practice than other temporal scales because most engineering applications require information on the variability of the daily precipitation time series. However, due to the intermittency of this process (wet and dry days), the daily precipitation time series is difficult to model. An alternative approach is to separate the discussion of the modeling of the daily precipitation time series into two components: the modeling of daily precipitation occurrences, and the modeling of daily precipitation amounts. This study focus on the second category, which is the modeling of daily precipitation amounts, i.e. the modeling of the precipitation amounts on wet days.

Early studies attempted to incorporate the information of precipitation occurrences while modeling precipitation amounts. As a result, these studies ended up with models that either focus on the modeling of the total precipitation amounts of given periods or simulate precipitation amounts conditional on the wet and dry occurrences of previous days (Istok and Boersma, 1989; Nguyen, 1984; Nguyen and Rousselle, 1981; Stern and Coe, 1984; Todorovic and Woolhiser, 1975). However, in both cases, the main difficulty in modeling precipitation amounts is the selection of an appropriate probability distribution, and the relationship between precipitation occurrences and precipitation amounts is relatively unimportant (Ison et al., 1971; Katz, 1977; Ozturk, 1981; Smith and Schreiber, 1974; Woolhiser and Roldan, 1982). Thus, it is reasonable to model precipitation amounts without the consideration of precipitation occurrences and view the precipitation amounts of wet days of a single site as observations that are statistically mutually independent (Chandler and Wheater, 2002; Richardson, 1981; Wilks, 1998). However, the selection of an appropriate probability distribution for the modeling of daily precipitation amounts is still an open question (Nguyen and Mayabi, 1990).

The exponential, Gamma, Weibull, and mixed exponential distributions are the commonly used probability distributions in modeling daily precipitation amounts (Nguyen and Mayabi, 1990). The exponential distribution is a special case of the Gamma/Weibull distribution with skewness equal to two. As a result, the Gamma and Weibull distributions can provide better fits to the observed daily precipitation amounts than the exponential distribution. However, a better fit does not necessarily ensure a better model because of the risk of the overfitting, and the exponential distribution is still an appropriate distribution for the modeling of daily precipitation amounts in certain situations. Furthermore, the appropriate probability distribution for the modeling of daily precipitation amounts falls between the Gamma and Weibull distributions; and the mixed exponential distribution is appropriate for modeling daily precipitation amounts in general because it serves as a probability distribution between the Gamma and Weibull distributions (Nguyen and Mayabi, 1990; Smith and Schreiber, 1974; Woolhiser and Roldan, 1982). However, unlike how the exponential distribution is a special case of the Gamma/Weibull distribution, the Gamma and Weibull distributions are not special cases of the mixed exponential distribution. There is no guarantee that the mixed exponential distribution is always a better distribution than the Gamma and Weibull distributions for the modeling of daily precipitation amounts. Based on the above, a unique probability distribution that provides the best model for the modeling of daily precipitation amounts in all situations does not exist. A framework that integrates all the commonly used distributions into one comparison base is needed. Hence, in this study the mixed Gamma Weibull distribution (MGW) was selected to serve this role (Everitt and Hand, 1981). The exponential, Gamma, Weibull, and mixed exponential distributions are all special cases of the MGW distribution. As a result, the MGW distribution integrates all the commonly used probability distributions into one framework. Finally, via the large sample inference of likelihood ratios, a model selection criterion can be established to identify the appropriate model for the modeling of daily precipitation amounts from the MGW distribution framework.

## Methodology

Different from the application of statistics to other fields, the modeling of precipitation amounts focuses on capturing not only the mean but also the extreme values of precipitation amounts. Therefore, one can say that a major part of hydrology research is the research on the convergence rate of tail probability of right skewed distribution functions. However, the commonly used Gamma and Weibull distributions limit themselves to having the convergence rates of tail probabilities, respectively, in the forms of $x^{\alpha-1}\exp(-x)$ and $x^{k-1}\exp(-x^k)$; and it has been recognized that the behavior of the heavy tail probability of daily precipitation amounts falls between these two forms of convergence rates (Nguyen and Mayabi, 1990). Thus, a probability distribution with a more flexible heavy tail probability behavior is needed for the modeling of daily precipitation amounts.

$$\begin{aligned} f_X(x \mid p,\alpha,\beta,k,\lambda) &= p g_1(x \mid \alpha,\beta) + (1-p) g_2(x \mid k,\lambda) \\ &= p\frac{1}{\Gamma(\alpha)\beta^{\alpha}} x^{\alpha-1}\exp\left(-\frac{x}{\beta}\right) + (1-p)\left(\frac{k}{\lambda}\right)\left(\frac{x}{\lambda}\right)^{k-1}\exp\left(-\left(\frac{x}{\lambda}\right)^{k}\right) \qquad (1) \\ &\text{with } 0 \le p \le 1,\ \ \alpha,\beta,k,\lambda > 0 \end{aligned}$$

Based on the above, the MGW distribution provides an appropriate framework for modeling of daily precipitation amounts. As shown in Equation (1), the probability density function (PDF) of the MGW distribution is defined as the linear weighting of the PDF of the Gamma and Weibull distributions. Therefore, the MGW distribution can have more flexibility in describing the heavy tail probability behavior of daily precipitation amounts since the tail probability of the MGW distribution is a mixture of two different forms of convergence rates of tail probabilities. In addition, the MGW distribution integrates the commonly used probability distributions, i.e. the exponential ($\alpha = k = 1$ and $\beta = \lambda$), Gamma ($p = 1$), Weibull ($p = 0$), and mixed exponential ($\alpha = k = 1$) distributions, into one framework. Finally, under the MGW distribution framework, the large sample inference of likelihood ratios can be applied to select the appropriate probability distributions for modeling daily precipitation amounts.

In this section, the commonly used probability distributions and the corresponding parameter estimation approaches will be first reviewed. Then, the parameter estimation methods of MGW distribution will follow. Finally, the likelihood ratio based model selection approach will be introduced to integrate all the above mentioned probability distributions into one comparison framework.

*The commonly used probability distributions*. The exponential, Gamma, Weibull, and mixed exponential distributions with the maximum likelihood (ML) estimation method are the commonly used modeling approach for daily precipitation amounts. With $\alpha = k = 1$ and $\beta = \lambda$, the ML estimator of the exponential distribution is the result of solving the score function set to equal to zero, and can be expressed as Equation (2). Similarly, with $p$ equals to one and zero respectively, the ML estimators of the Gamma and Weibull distributions can be solved by setting the score functions equal to zero. Therefore, based on Equation (1), the ML estimators of the Gamma and Weibull distributions are the solutions of Equation (3) and Equation (4) respectively and can be solved numerically.

$$\beta = \lambda = \frac{1}{n}\sum_{i=1}^{n} x_i \tag{2}$$

$$S(\alpha, \beta \mid \tilde{X}) = \begin{pmatrix} \dfrac{\partial l(\alpha, \beta \mid \tilde{X})}{\partial \alpha} \\ \dfrac{\partial l(\alpha, \beta \mid \tilde{X})}{\partial \beta} \end{pmatrix} = \begin{pmatrix} -n\psi(\alpha) - n\ln\beta + \sum_{i=1}^{n} \ln x_i \\ -n\dfrac{\alpha}{\beta} + \dfrac{1}{\beta^2}\sum_{i=1}^{n} x_i \end{pmatrix} \overset{set}{=} 0 \tag{3}$$

$$S(k, \lambda \mid \tilde{X}) = \begin{pmatrix} \dfrac{\partial l(k, \lambda \mid \tilde{X})}{\partial k} \\ \dfrac{\partial l(k, \lambda \mid \tilde{X})}{\partial \lambda} \end{pmatrix} = \begin{pmatrix} \dfrac{n}{k} - n\ln\lambda + \sum_{i=1}^{n}\left[\ln x_i - \left(\dfrac{x_i}{\lambda}\right)^k \ln\left(\dfrac{x_i}{\lambda}\right)\right] \\ \dfrac{k}{\lambda}\sum_{i=1}^{n}\left(\dfrac{x_i}{\lambda}\right)^k - n\dfrac{k}{\lambda} \end{pmatrix} \overset{set}{=} 0 \tag{4}$$

where $\tilde{X} = (X_1, \ldots, X_n)^T$ represents the random sample, $S(\cdot \mid \tilde{X})$ represents the score function of the random sample $\tilde{X}$, $l(\cdot \mid \tilde{X})$ represents the logarithm of the likelihood function $L(\cdot \mid \tilde{X})$ of the random sample $\tilde{X}$, and $\psi(\cdot)$ represents the digamma function.

The ML estimator of the mixed exponential distribution can be obtained by adopting the monotonic property of the expectation and maximization (EM) series (Dempster et al., 1977). Assume that the random sample $\tilde{X}$ is just the observable part of the full data, and the random sample $\tilde{Z}$ represents the unobservable part of the full data. Then, it can be proved that the EM series $\hat{\theta}_{(r)}$ obtained from Equation (5) has the monotonic property of Equation (6).

$$\hat{\theta}_{(r+1)} = \arg\max_{\theta} E\left(\ln L(\theta \mid \tilde{X}, \tilde{Z}) \mid \hat{\theta}_{(r)}, \tilde{X}\right) \tag{5}$$

$$L(\hat{\theta}_{(r+1)} \mid \tilde{X}) \geq L(\hat{\theta}_{(r)} \mid \tilde{X}) \tag{6}$$

where $\theta$ represents the parameter vector of the model, $L(\theta \mid \tilde{X}, \tilde{Z})$ represents the likelihood function of the full data $(\tilde{X}, \tilde{Z})$, $L(\theta \mid \tilde{X})$ represents the likelihood function of the observable data $\tilde{X}$, and $E\left(\cdot \mid \hat{\theta}_{(r)}, \tilde{X}\right)$ is the expectation conditional on the observable data $\tilde{X}$ and uses $\hat{\theta}_{(r)}$ as the value of $\theta$ when taking the expectation. The monotonic property of Equation (6) provides a solution for finding ML estimators when the likelihood function of the observed data $L(\theta \mid \tilde{X})$ is more difficult to obtain than the likelihood function of the full data $L(\theta \mid \tilde{X}, \tilde{Z})$.

The realizations of mixture distributions can be viewed as the result of the hierarchy simulation that first simulates the belonging probability distribution within the mixture structure of the realization to be simulated by using the mixture weights. Then, this hierarchy simulation simulates the realization by using the corresponding probability distribution. Therefore, by taking $\tilde{Z}$ as the random sample of the simulation of the first step, the monotonic property of EM series can be applied to find the ML estimators of mixture distributions. For mixture distributions with only two mixture components, which is the case of Equation (1), $\tilde{Z} = (Z_1, \ldots, Z_n)^T$ is a random sample of the Bernoulli trial. Thus, for the mixed exponential distribution case, the expectation of Equation (5) can be written as the following.

$$\begin{aligned}
&\sum_{i=1}^{n} E\left[ \ln\left[Z_i p + (1-Z_i)(1-p)\right] + \ln\left( \frac{Z_i}{\beta}\exp\left(-\frac{X_i}{\beta}\right) + \frac{(1-Z_i)}{\lambda}\exp\left(-\frac{X_i}{\lambda}\right)\right) \middle| \hat{p}_{(r)}, \hat{\beta}_{(r)}, \hat{\lambda}_{(r)}, \tilde{X} \right] \\
&= \sum_{i=1}^{n}\left[ \hat{p}_{ME|x_i}^{(r)}\left(\ln p - \ln\beta - \frac{x_i}{\beta}\right) + \left(1 - \hat{p}_{ME|x_i}^{(r)}\right)\left(\ln(1-p) - \ln\lambda - \frac{x_i}{\lambda}\right)\right]
\end{aligned} \tag{7}$$

with $$\hat{p}_{ME|x_i}^{(r)} = \frac{\dfrac{\hat{p}_{(r)}}{\hat{\beta}_{(r)}}\exp\left(-\dfrac{x_i}{\hat{\beta}_{(r)}}\right)}{\dfrac{\hat{p}_{(r)}}{\hat{\beta}_{(r)}}\exp\left(-\dfrac{x_i}{\hat{\beta}_{(r)}}\right) + \dfrac{(1-\hat{p}_{(r)})}{\hat{\lambda}_{(r)}}\exp\left(-\dfrac{x_i}{\hat{\lambda}_{(r)}}\right)}$$

Then, $\hat{p}_{(r+1)}$, $\hat{\beta}_{(r+1)}$, and $\hat{\lambda}_{(r+1)}$ can be obtained by solving the first order derivatives of Equation (7) with respect to $p$, $\beta$, and $\lambda$ equal to zero and can be expressed as the following.

$$
\begin{aligned}
\hat{p}_{(r+1)} &= \frac{1}{n}\sum_{i=1}^{n} \hat{p}_{ME|x_i}^{(r)} \\
\hat{\beta}_{(r+1)} &= \frac{1}{n\hat{p}_{(r+1)}}\sum_{i=1}^{n} x_i \hat{p}_{ME|x_i}^{(r)} \\
\hat{\lambda}_{(r+1)} &= \frac{1}{n(1-\hat{p}_{(r+1)})}\sum_{i=1}^{n} x_i \left(1-\hat{p}_{ME|x_i}^{(r)}\right)
\end{aligned}
\qquad (8)
$$

Finally, the ML estimator of the mixed exponential distribution can be calculated iteratively by using Equation (8).

*The mixed Gamma Weibull distribution with the mixture estimation approach*. The MGW distribution is a very flexible distribution, and the shape of the PDF of the MGW distribution is not necessarily unimodal or monotonically decreasing. Therefore, if an unsuitable estimation method is applied, the usage of the MGW distribution in the modeling of daily precipitation amounts may result in unrealistic results. Hence, a parameter estimation approach of the MGW distribution for the modeling of daily precipitation amounts is needed. The mixture estimation approach provides a solution to this problem by combining both the method of moment (MOM) estimation and the ML estimation approaches.

For hydrological purposes, it is appreciated that the daily precipitation amount model can capture the mean, variance, and extremes of the observed daily precipitation amounts. Thus, the mixture estimation approach aims at searching for the ML estimator at the restricted parameter space while forcing the mean and variance of the model equal to the observed mean and variance (MOM estimation). This estimation criterion satisfies the expectation of the daily precipitation amount modeling from the hydrological point of view.

By taking the first and second order moments of the MGW random variable $X$ of Equation (1), the following equations for moments can be obtained.

$$
\begin{aligned}
p\alpha\beta_* + (1-p)\lambda_*\Gamma\left(1+\frac{1}{k}\right) &= 1 \\
p\alpha(\alpha+1)\beta_*^2 + (1-p)\lambda_*^2\Gamma\left(1+\frac{2}{k}\right) &= 1+CV
\end{aligned}
\qquad (9)
$$

with $\beta_* = \dfrac{\beta}{E(X)}$, $\lambda_* = \dfrac{\lambda}{E(X)}$, and $CV = \dfrac{Var(X)}{[E(X)]^2}$

By combining the first and second lines of Equation (9), Equation (10) can be obtained.

$$\left[p\alpha(\alpha+1) + \frac{p^2\alpha^2}{(1-p)} \frac{\Gamma\left(1+\frac{2}{k}\right)}{\left[\Gamma\left(1+\frac{1}{k}\right)\right]^2}\right]\beta_*^2 + \left[-2\frac{p\alpha}{(1-p)} \frac{\Gamma\left(1+\frac{2}{k}\right)}{\left[\Gamma\left(1+\frac{1}{k}\right)\right]^2}\right]\beta_* + \left[\frac{1}{(1-p)} \frac{\Gamma\left(1+\frac{2}{k}\right)}{\left[\Gamma\left(1+\frac{1}{k}\right)\right]^2} - 1 - CV\right] = 0 \tag{10}$$

Therefore, $\beta_*$ can simply be solved as a quadratic equation if the values of $p$, $\alpha$, $k$, and $CV$ are given; and $\lambda_*$ can be obtained from the first line of Equation (9) when $\beta_*$ is obtained. Furthermore, Equation (10) has exactly one and two positive solution(s) for $\beta_*$ when, respectively, Equation (11) and Equation (12) are satisfied.

$$CV \geq \frac{1}{(1-p)} \frac{\Gamma\left(1+\frac{2}{k}\right)}{\left[\Gamma\left(1+\frac{1}{k}\right)\right]^2} - 1 \tag{11}$$

$$\begin{aligned} &\frac{1}{(1-p)} \frac{\Gamma\left(1+\frac{2}{k}\right)}{\left[\Gamma\left(1+\frac{1}{k}\right)\right]^2} - 1 > CV \geq \\ &\frac{1}{(1-p)} \frac{\Gamma\left(1+\frac{2}{k}\right)}{\left[\Gamma\left(1+\frac{1}{k}\right)\right]^2} - \left[\frac{p\alpha}{(1-p)} \frac{\Gamma\left(1+\frac{2}{k}\right)}{\left[\Gamma\left(1+\frac{1}{k}\right)\right]^2}\right]^2 \Bigg/ \left[p\alpha(\alpha+1) + \frac{p^2\alpha^2}{(1-p)} \frac{\Gamma\left(1+\frac{2}{k}\right)}{\left[\Gamma\left(1+\frac{1}{k}\right)\right]^2}\right] - 1 \end{aligned} \tag{12}$$

Finally, based on the first line of Equation (9), to have positive $\lambda_*$, Equation (13) needs to be satisfied.

$$\beta_* < \frac{1}{p\alpha} \tag{13}$$

Therefore, the mixture estimation of the MGW distribution can be done by first substituting the sample mean and the sample variance into $CV$. Then, a mesh grid of different combinations of values of $p$, $\alpha$, and $k$ between zero and one is constructed, and the corresponding $\beta$ and $\lambda$ of each grid are solved by applying Equations (9) to (13) and substituting $E(X)$ with the sample mean. Finally, the mixture estimator can be obtained by identifying the parameter combination that has the maximum likelihood among all other parameter combinations.

Because the construction of the mesh grid of $p$, $\alpha$, and $k$ is independent of data, the usage of Equations (11) to (13) can save the computation by screening out the combinations of $(p, \alpha, k)$ that don't have the corresponding positive solution(s) of $\beta$ and $\lambda$. In addition, the range restriction of $\alpha$ and $k$ during the construction of the mesh grid ensures that the PDF of the MGW distribution with the mixture estimation is monotonically decreasing, which is the case with the majority of the observations of daily precipitation amounts. Finally, the interval of the grids of $\alpha$ and $k$ can use the corresponding skewness of Gamma $\gamma_\alpha$ and skewness of Weibull $\gamma_k$ as references.

Based on the above discussion, the basic principle for the proposed estimation method is to use $\beta$ and $\lambda$ for adjusting the PDF of the MGW distribution to have the same scale as the data; and the parameters $p$, $\alpha$, and $k$ play the role in controlling the tail probability of the PDF of the MGW distribution to fit the data.

*The likelihood ratio based model selection criterion.* If the ML estimators found at both the low and high dimensional parameter spaces are consistent estimators of the true parameter, and the score functions of both the ML estimators are zero vectors, then, under the regularity conditions, negative two times the logarithm of the likelihood ratio of these two ML estimators has a limiting chi-square distribution with a degree of freedom equal to the dimension difference (Shao, 2003). Therefore, based on this chi-square approximation property, limiting size $\alpha$ tests can be constructed for model selection purposes. However, the location of the true parameter actually depends on the assumption of the parameter space, which should also take the physical phenomenon of daily precipitation amounts into consideration. Thus, the following section will first focus on the searching of candidate ML estimators that have zero score functions for the MGW distribution. Then, the ML estimator of the MGW distribution for the modeling of daily precipitation amounts can be identified from the candidate ML estimators.

Similar to the ML estimation of the mixed exponential distribution case, the monotonic property of the EM series can be applied here for the searching of the candidate ML estimators for the MGW distribution. However, it is difficult to control to which candidate ML estimator the direct application of the EM series converges, though the EM series guarantees the monotonic property of Equation (6). As a result, a modified EM algorithm is introduced here by combining the EM algorithm with the gradient descent approach for the searching of the candidate ML estimators. The iteration process first uses the EM algorithm to find the values of $p$, $\beta$, and $\lambda$ that maximize the likelihood function when the values of $\alpha$ and $k$ are given. Then, the score function at the parameter location of the current iteration is evaluated. Finally, the values of $\alpha$ and $k$ are updated by using the score function with the gradient descent approach.

When the values of $\alpha$ and $k$ are given as $\hat{\alpha}^{(m)}$ and $\hat{k}^{(m)}$ respectively, the MGW distribution can be viewed as a $p$, $\beta$, and $\lambda$ three-parameter distribution. Then, the expectation of Equation (5) can be written as the following.

$$\sum_{i=1}^{n} E\left[\begin{matrix}\ln\left[Z_i p + (1-Z_i)(1-p)\right] \\ +\ln\left(Z_i g_1(X_i \mid \hat{\alpha}^{(m)}, \beta) + (1-Z_i) g_2(X_i \mid \hat{k}^{(m)}, \lambda)\right)\end{matrix}\middle| \hat{p}_{(r)}^{(m)}, \hat{\beta}_{(r)}^{(m)}, \hat{\lambda}_{(r)}^{(m)}, \tilde{X}\right]$$

$$= \sum_{i=1}^{n}\left[\hat{p}_{MGW|x_i}^{(m,r)}\left(\ln p + \ln g_1(x_i \mid \hat{\alpha}^{(m)}, \beta)\right) + \left(1-\hat{p}_{MGW|x_i}^{(m,r)}\right)\left(\ln(1-p) + \ln g_2(x_i \mid \hat{k}^{(m)}, \lambda)\right)\right] \tag{14}$$

with $\hat{p}_{MGW|x_i}^{(m,r)} = \dfrac{\hat{p}_{(r)}^{(m)} g_1(x_i \mid \hat{\alpha}^{(m)}, \beta_{(r)}^{(m)})}{f_X(x \mid \hat{p}_{(r)}^{(m)}, \hat{\alpha}^{(m)}, \hat{\beta}_{(r)}^{(m)}, \hat{k}^{(m)}, \hat{\lambda}_{(r)}^{(m)})}$

Then, $\hat{p}_{(r+1)}^{(m)}$, $\hat{\beta}_{(r+1)}^{(m)}$, and $\hat{\lambda}_{(r+1)}^{(m)}$ can be obtained by using the same fashion as the mixed exponential distribution case and can be expressed as the following.

$$\begin{aligned}
\hat{p}_{(r+1)}^{(m)} &= \frac{1}{n}\sum_{i=1}^{n}\hat{p}_{MGW|x_i}^{(m,r)} \\
\hat{\beta}_{(r+1)}^{(m)} &= \frac{1}{n\hat{\alpha}^{(m)}\hat{p}_{(r+1)}^{(m)}}\sum_{i=1}^{n} x_i \hat{p}_{MGW|x_i}^{(m,r)} \\
\hat{\lambda}_{(r+1)}^{(m)} &= \left[\frac{1}{n(1-\hat{p}_{(r+1)}^{(m)})}\sum_{i=1}^{n} x_i^{\hat{k}^{(m)}}\left(1-\hat{p}_{MGW|x_i}^{(m,r)}\right)\right]^{\frac{1}{\hat{k}^{(m)}}}
\end{aligned} \tag{15}$$

Finally, $\hat{\alpha}^{(m+1)}$ and $\hat{k}^{(m+1)}$ can be obtained using the gradient descent approach indicated by Equation (16) after the convergence values $\hat{p}^{(m)}$, $\hat{\beta}^{(m)}$, and $\hat{\lambda}^{(m)}$ are obtained via the iteration procedure of Equation (15).

$$\begin{aligned}
\hat{\alpha}^{(m+1)} &= \hat{\alpha}^{(m)} + \min(\varepsilon, 1)\left.\frac{\partial l(p, \alpha, \beta, k, \lambda \mid \tilde{X})}{\partial \alpha}\right|_{(p,\alpha,\beta,k,\lambda)=(\hat{p}^{(m)}, \hat{\alpha}^{(m)}, \hat{\beta},^{(m)} \hat{k}^{(m)}, \hat{\lambda}^{(m)})} \\
\hat{k}^{(m+1)} &= \hat{k}^{(m)} + \min(\varepsilon, 1)\left.\frac{\partial l(p, \alpha, \beta, k, \lambda \mid \tilde{X})}{\partial k}\right|_{(p,\alpha,\beta,k,\lambda)=(\hat{p}^{(m)}, \hat{\alpha}^{(m)}, \hat{\beta},^{(m)} \hat{k}^{(m)}, \hat{\lambda}^{(m)})}
\end{aligned} \tag{16}$$

where $\varepsilon$ is a positive constant used for controlling the convergence speed of the gradient descent algorithm. The minimum operator forces this term not to exceed one. Furthermore, to automatically adjust the convergence speed of the gradient descent algorithm, the value of $\varepsilon$ is calculated by using Equation (17).

$$\varepsilon = \varepsilon_0 \left\| \left. \frac{\begin{pmatrix} \frac{\partial l(p,\alpha,\beta,k,\lambda|\tilde{X})}{\partial \alpha} \\ \frac{\partial l(p,\alpha,\beta,k,\lambda|\tilde{X})}{\partial k} \end{pmatrix}^T \begin{pmatrix} \frac{\partial l(p,\alpha,\beta,k,\lambda|\tilde{X})}{\partial \alpha} \\ \frac{\partial l(p,\alpha,\beta,k,\lambda|\tilde{X})}{\partial k} \end{pmatrix}}{\begin{pmatrix} \frac{\partial l(p,\alpha,\beta,k,\lambda|\tilde{X})}{\partial \alpha} \\ \frac{\partial l(p,\alpha,\beta,k,\lambda|\tilde{X})}{\partial k} \end{pmatrix}^T \begin{pmatrix} \frac{\partial^2 l(p,\alpha,\beta,k,\lambda|\tilde{X})}{\partial \alpha^2} & \frac{\partial^2 l(p,\alpha,\beta,k,\lambda|\tilde{X})}{\partial \alpha \partial k} \\ \frac{\partial^2 l(p,\alpha,\beta,k,\lambda|\tilde{X})}{\partial k \partial \alpha} & \frac{\partial^2 l(p,\alpha,\beta,k,\lambda|\tilde{X})}{\partial k^2} \end{pmatrix} \begin{pmatrix} \frac{\partial l(p,\alpha,\beta,k,\lambda|\tilde{X})}{\partial \alpha} \\ \frac{\partial l(p,\alpha,\beta,k,\lambda|\tilde{X})}{\partial k} \end{pmatrix}} \right|_{(p,\alpha,\beta,k,\lambda)=(\hat{p}^{(m)},\hat{\alpha}^{(m)},\hat{\beta}^{(m)},\hat{k}^{(m)},\hat{\lambda}^{(m)})} \right\|$$

with $$\frac{\partial^2 l(p,\alpha,\beta,k,\lambda|\tilde{X})}{\partial \alpha^2} = \sum_{i=1}^{n} \frac{p g_1(x_i|\alpha,\beta)}{f_X(x_i|p,\alpha,\beta,k,\lambda)} \left[ (\ln x_i - \psi(\alpha) - \ln\beta)^2 - \frac{d\psi(\alpha)}{d\alpha} \right] - \sum_{i=1}^{n} \left[ \frac{p g_1(x_i|\alpha,\beta)(\ln x_i - \psi(\alpha) - \ln\beta)}{f_X(x_i|p,\alpha,\beta,k,\lambda)} \right]^2 \quad (17)$$

$$\frac{\partial^2 l(p,\alpha,\beta,k,\lambda|\tilde{X})}{\partial k^2} = \sum_{i=1}^{n} \left[ \begin{array}{l} \frac{(1-p) g_2(x|k,\lambda)}{f_X(x_i|p,\alpha,\beta,k,\lambda)} \left[ \left( \frac{1}{k} + \ln\left(\frac{x_i}{\lambda}\right) - \left(\frac{x_i}{\lambda}\right)^k \ln\left(\frac{x_i}{\lambda}\right) \right)^2 - \frac{1}{k^2} - \left( \ln\left(\frac{x_i}{\lambda}\right) \right)^2 \left(\frac{x_i}{\lambda}\right)^k \right] \\ - \left[ \frac{(1-p) g_2(x_i|k,\lambda)}{f_X(x_i|p,\alpha,\beta,k,\lambda)} \left( \frac{1}{k} + \ln\left(\frac{x_i}{\lambda}\right) - \left(\frac{x_i}{\lambda}\right)^k \ln\left(\frac{x_i}{\lambda}\right) \right) \right]^2 \end{array} \right]$$

$$\frac{\partial^2 l(p,\alpha,\beta,k,\lambda|\tilde{X})}{\partial \alpha \partial k} = \frac{\partial^2 l(p,\alpha,\beta,k,\lambda|\tilde{X})}{\partial k \partial \alpha} = -\sum_{i=1}^{n} \frac{p g_1(x_i|\alpha,\beta)(\ln x_i - \psi(\alpha) - \ln\beta)}{f_X(x_i|p,\alpha,\beta,k,\lambda)} \frac{(1-p) g_2(x_i|k,\lambda)}{f_X(x_i|p,\alpha,\beta,k,\lambda)} \left( \frac{1}{k} + \ln\left(\frac{x_i}{\lambda}\right) - \left(\frac{x_i}{\lambda}\right)^k \ln\left(\frac{x_i}{\lambda}\right) \right)$$

Equation (17) is the result of substituting Equation (16) (replacing the minimum operator by $\varepsilon$) into the second order Taylor expansion of the log-likelihood function of the MGW distribution with respect to $(\alpha, k)^T$ and forcing the absolute value of the ratio of the second order term over the first order term to be the positive constant $\varepsilon_0$. Therefore, by using Equation (17), the iteration speed can be automatically controlled by using the second order derivative as a reference. By controlling $\varepsilon_0$, one can ensure that the iteration step indicated by Equation (16) still obeys the first order Taylor expansion approximation. However, when the second order term is equal or very close to zero, the adoption of Equation (17) is meaningless. In this situation, the value of $\varepsilon$ is replaced by the value one as indicated by the minimum operator of Equation (16).

Finally, the iteration stops when a candidate ML estimator that has a zero score function, expressed as Equation (18), is found. Alternatively, the first element within Equation (18) is neglected when $p$ is very close to one or zero because these two situations indicate that the MGW distribution has degenerated to the Gamma or Weibull distribution.

$$S(p,\alpha,\beta,k,\lambda \,|\, \tilde{X}) = \begin{pmatrix} \frac{\partial l(p,\alpha,\beta,k,\lambda|\tilde{X})}{\partial p} \\ \frac{\partial l(p,\alpha,\beta,k,\lambda|\tilde{X})}{\partial \alpha} \\ \frac{\partial l(p,\alpha,\beta,k,\lambda|\tilde{X})}{\partial \beta} \\ \frac{\partial l(p,\alpha,\beta,k,\lambda|\tilde{X})}{\partial k} \\ \frac{\partial l(p,\alpha,\beta,k,\lambda|\tilde{X})}{\partial \lambda} \end{pmatrix} = \sum_{i=1}^{n} \begin{pmatrix} \frac{g_1(x_i|\alpha,\beta) - g_2(x_i|k,\lambda)}{f_X(x_i|p,\alpha,\beta,k,\lambda)} \\ \frac{p(\ln x_i - \psi(\alpha) - \ln\beta) g_1(x_i|\alpha,\beta)}{f_X(x_i|p,\alpha,\beta,k,\lambda)} \\ \frac{p(x_i - \alpha\beta) g_1(x_i|\alpha,\beta)}{\beta^2 f_X(x_i|p,\alpha,\beta,k,\lambda)} \\ \frac{(1-p)\left( \frac{1}{k} + \ln\left(\frac{x_i}{\lambda}\right) - \left(\frac{x_i}{\lambda}\right)^k \ln\left(\frac{x_i}{\lambda}\right) \right) g_2(x_i|k,\lambda)}{f_X(x_i|p,\alpha,\beta,k,\lambda)} \\ \frac{(1-p)\left( \left(\frac{k}{\lambda}\right)\left(\frac{x_i}{\lambda}\right)^k - \frac{k}{\lambda} \right) g_2(x_i|k,\lambda)}{f_X(x_i|p,\alpha,\beta,k,\lambda)} \end{pmatrix} \quad (18)$$

With different initial values of the parameter at the beginning of the algorithm, the iteration process of Equations (15) to (17) ends up with different candidate ML estimators. Then, the ML estimator of the MGW distribution for the modeling of daily precipitation amounts can be obtained by selecting the candidate ML estimator that has the maximum likelihood among all candidate ML estimators that have a unimodal or monotonically decreasing PDF of the MGW distribution.

The diagnosis of the shape of the PDF of the MGW distribution can simply be done by numerically checking the first order derivative of Equation (1) with respect to $x$. The diagnosis of the shape of the PDF of the MGW distribution defines the assumption of the parameter space and the location of the true parameter for the application of the MGW distribution to the modeling of daily precipitation amounts. In other words, the parameter space is defined at the area that the corresponding PDF of the MGW distribution is unimodal or monotonically decreasing because PDFs with shapes other than these two shapes are not physically realistic when considering the modeling of daily precipitation amounts. Therefore, the ML estimator of this defined parameter space is a consistent estimator of the true parameter under this assumption of the parameter space. In addition, the MGW distribution with the parameter space under this definition still covers the exponential, Gamma, Weibull, mixed exponential, and MGW (mixture estimation) distributions into one framework. Finally, based on the above discussions, the model selection of the daily precipitation amount modeling can be constructed by treating the likelihood of the ML estimator of the MGW distribution as a comparative base and using the chi-square approximation property of likelihood ratios.

## Study Area

In this study, the modeling approaches and the model selection method are applied to the daily precipitation record of southern Quebec as an illustrative example. The spatial distribution of the raingauges within the study area is described by Chen (2020). The temporal coverage of the daily precipitation records of all the study sites ranges uniformly from 1961-Jan-01 to 2001-Dec-31, and the daily precipitation data is separated into two periods: the calibration period and the validation period. The calibration and validation periods represents, respectively, the period 1961-Jan-01 to 1985-Dec-31 and the period 1986-Jan-01 to 2001-Dec-31. However, the modeling approaches and the model selection method discussed in this study only involve the model fitting of daily precipitation amounts, i.e. model calibration. Thus, only the observations of daily precipitation within the calibration period are adopted in this study while the observations of daily precipitation within the validation period are reserved for future studies.

## Results and Discussion

In this study, a day is considered a wet day if the accumulated precipitation of that day is greater than or equal to $1\,mm$ while a day is considered a dry day if the accumulated precipitation of that day is less than $1\,mm$. More specifically, in this study, the modeling of daily precipitation amounts means the modeling of the daily precipitation amounts of wet days. However, under this definition, the observations of the daily precipitation amounts are always greater than or equal to $1\,mm$ while the support of the PDFs of the exponential, Gamma, Weibull, mixed exponential, and MGW distribution is $(0, \infty)$. Therefore, the direct modeling of the observations of the daily precipitation amounts is not appropriate due to this inconsistency of the minimum value.

Although the minimum value of the observations of the daily precipitation amounts is $1\,mm$, it actually means that those $1\,mm$ values are the result of rounding the daily accumulated precipitation that ranged from $0.95\,mm$ to $1.05\,mm$ because the raingauge observations have a $0.1\,mm$ increment. Thus, an offset value $0.95\,mm$ is subtracted from the observations of the daily precipitation amounts before modeling so that the minimum value of the input data is consistent with the support of the PDFs of the candidate distributions. Finally, the $0.95\,mm$ offset value is added back to the fitted models as a location parameter so that the fitted models can have a support of $(1\,mm, \infty)$ after rounding.

Furthermore, to account for seasonality, each candidate distribution is fitted to the observations of the daily precipitation amounts of each study site separately in each calendar month. For the ML estimation of the exponential, Gamma, and Weibull distributions, the calculations can simply be done by the direct application of Equations (2), (3), and (4) respectively. For the ML estimation of the mixed exponential distribution, the ML estimators can be obtained by adopting the iteration process indicated by Equation (8) until convergence. In this study, the initial values of $p$, $\beta$, and $\lambda$ of the iteration process use 0.5, $\frac{1.6}{n}\sum_{i=1}^{n} x_i$, and $\frac{0.4}{n}\sum_{i=1}^{n} x_i$ respectively. In addition, when $\beta = \lambda$, the ML estimation of the mixed exponential distribution is equivalent to the ML estimation of the exponential distribution. For the mixture estimation of the MGW distribution, the estimators can be calculated by applying Equations (9) to (13). The grids of $p$ ranges from 0 to 1 with a uniform increment 0.01, and the grids of both $\gamma_\alpha$ and $\gamma_k$ ranges from 2 to 5 with a uniform increment 0.01. When the value of $CV$ calculated from the sample is smaller than one, the mixture estimation of the MGW distribution cannot be applied since the simultaneous fitting of the mean and variance of the model to the observation is not possible. When the estimated $p$ is equal to one and zero respectively, the mixture estimation of the MGW distribution cannot be applied to the model selection of the likelihood ratios because the mixture estimation gives the MOM estimation result of the Gamma and Weibull distribution. When the estimated $\alpha$ and $k$ are equal to one respectively, the mixture estimation of the MGW distribution is equivalent to the mixture estimation of the mixed exponential Weibull (MEW) and the mixed Gamma exponential (MGE) distribution. When both the estimated $\alpha$ and $k$ are equal to one, the mixture estimation of the MGW distribution is equivalent to the mixture estimation of the mixed exponential distribution.

The ML estimation of the MGW distribution can be achieved by following the iteration processes indicated by Equations (15) to (18). The iteration process of Equation (15) is stopped and moved to the iteration process of Equation (16) when the absolute values of the first, third, and fifth elements within Equation (18) are smaller than $10^{-3}$, and the positive constant $\varepsilon_0$ within Equation (17) adopts the value $0.01$. Finally, the whole iteration process is stopped when the absolute values of all the elements of the score function indicated by Equation (18) are smaller than $10^{-3}$. When the first element of Equation (18) is positive with $p > 1-10^{-4}$ and negative with $p < 10^{-4}$ respectively, the MGW distribution is deemed that it has degenerated to the Gamma and Weibull distributions. As a result, in these two situations, the first element of Equation (18) is neglect when determining whether to stop the iteration processes of Equations (15) and (16) or not. By using the above settings, a stable convergence of the iteration process with a satisfactory accuracy of the final result is obtained.

Twelve sets of initial parameters are used to begin the iteration process. The first set uses the ML estimator of the mixed exponential distribution with $\alpha = k = 1$ as the initial parameter. The second set switches the values of $\beta$ and $\lambda$ of the ML estimator of the mixed exponential distribution and changes the corresponding value of $p$ to the value $1-p$ as the initial parameter because the mixed exponential distribution has a symmetric parameter structure toward $\beta$ and $\lambda$. In the case that the ML estimation of the mixed exponential distribution results in the ML estimation of the exponential distribution, the first two sets of initial parameters are not adopted. The third set uses $p = 0.5$ and sets the ML estimators of the Gamma and the Weibull distributions as the values of $(\alpha, \beta)$ and $(k, \lambda)$ respectively. The fourth set uses the result of the mixture estimation of the MGW distribution as the initial parameter. The remaining eight sets of initial parameters consider the nine combinations of the values of $\gamma_\alpha$ and $\gamma_k$ equal to 1.5, 2, and 2.5 but exclude the combination that both the values of $\gamma_\alpha$ and $\gamma_k$ are equal to 2; and the values of $p$, $\beta$, and $\lambda$ adopts 0.5, $\frac{1}{\alpha\, n}\sum_{i=1}^{n} x_i$, and $\frac{1}{\Gamma(1+1/k)\, n}\sum_{i=1}^{n} x_i$ respectively. Furthermore, to save computation resources, a candidate ML estimator is removed from the candidate pool before the completion of the iteration when $\hat{\alpha}^{(m)}$ is greater than 25 with a corresponding variance of Gamma smaller than 0.01 or when $\hat{k}^{(m)}$ is greater than 25 with a corresponding variance of Weibull smaller than 0.01. It is because these two situations indicate that one of the two mixture components of the MGW distribution fits to a single value of the data and will have a convergence result that is not realistic when considering the modeling of daily precipitation amounts. Finally, the ML estimator of the MGW distribution can be selected from these candidate ML estimators after the completion of the iteration by checking the shapes of the PDFs of the MGW distribution.

However, there is no guarantee that all the candidate ML estimators can be found by using the above mentioned initial parameter settings with the iteration processes given by Equations (15) to (17) because the only way to guarantee that all the candidate ML estimators can be found is to examine all the possible values of the parameter vector. Nonetheless, the candidate ML estimators found by the above mentioned approach still represent the local optimums of the majority of the parameter space because the initial parameter settings of these candidate ML estimators represent the points of interest in the parameter space in our discussion. Thus, the ML estimator selected from the candidate ML estimators found by the above mentioned approach can be used as the comparative base of the likelihood ratio model selection approach. Finally, the values of the logarithm of the ML of all the candidate distributions are summarized in Table 1.

| Month | Site | Exponential | Gamma | Weibull | Mixed Exponential | MGW (mixture estimation) | MGW (ML estimation) |
|---|---|---|---|---|---|---|---|
| Jan | Dorval | -666.901 | -647.614 | -645.278 | -644.819 | -643.789(b1) | -642.260 |
| Jan | Cornwall | -730.334 | -726.343 | -724.875 | -722.121 | -721.970(b1) | -720.863 |
| Jan | Drummondville | -774.112 | -770.467 | -769.661 | -769.055 | -769.083(b2) | -769.661(c) |
| Jan | Farnham | -812.074 | -796.714 | -798.559 | -786.408 | -786.686(b2) | -796.179 |
| Jan | Lennoxville | -759.591 | -751.836 | -749.716 | -746.776 | -746.811(b3) | -746.328 |
| Jan | Morrisburg | -738.489 | -725.931 | -723.363 | -724.779 | -722.317 | -722.587 |
| Jan | Oka | -760.927 | -755.277 | -754.267 | -754.338 | -753.204(b1) | -752.129 |
| Jan | Ottawa CDA | -605.898 | -594.570 | -593.181 | -594.976 | -592.363(b1) | -591.002 |
| Jan | St. Alban | -636.514 | -635.504 | -634.763 | -633.020 | -633.045(b3) | -632.137 |
| Jan | St. Jerome | -713.249 | -707.937 | -705.757 | -701.399 | -701.533(b3) | -699.807 |
| Feb | Dorval | -591.983 | -581.718 | -581.371 | -581.028 | -580.673(b2) | -577.738 |
| Feb | Cornwall | -629.961 | -626.502 | -626.756 | -625.625 | -625.655(b1) | -625.569 |
| Feb | Drummondville | -622.320 | -615.726 | -614.719 | -615.483 | -613.094(b1) | -611.935 |
| Feb | Farnham | -738.064 | -714.446 | -717.088 | -705.716 | -705.058(b2) | -710.924 |
| Feb | Lennoxville | -638.105 | -623.400 | -624.176 | -621.298 | -621.187(b2) | -621.131 |
| Feb | Morrisburg | -655.265 | -651.996 | -651.363 | -651.038 | -650.813(b1) | -650.336 |
| Feb | Oka | -632.037 | -631.331 | -631.406 | -631.123 | -631.391(b2) | -630.500 |
| Feb | Ottawa CDA | -577.520 | -561.394 | -559.413 | -559.487 | -557.799(b2) | -555.952 |
| Feb | St. Alban | -594.294 | -585.409 | -585.205 | -583.830 | -583.823 | -583.804 |
| Feb | St. Jerome | -637.102 | -633.319 | -632.490 | -631.353 | -631.402(b3) | -630.787 |
| Mar | Dorval | -659.667 | -652.469 | -651.783 | -651.059 | -651.118(b1) | -651.783(c) |
| Mar | Cornwall | -579.294 | -572.783 | -572.782 | -573.492 | -572.194(b2) | -572.330 |
| Mar | Drummondville | -686.926 | -680.058 | -679.508 | -679.531 | -679.053(b2) | -675.558 |
| Mar | Farnham | -781.348 | -777.319 | -778.419 | -773.054 | -777.008(b2) | -776.115 |
| Mar | Lennoxville | -719.601 | -714.803 | -714.923 | -713.943 | -714.003(b1) | -714.066 |
| Mar | Morrisburg | -655.768 | -653.435 | -654.152 | -650.533 | -654.352(b2) | -652.231 |
| Mar | Oka | -687.278 | -682.990 | -683.435 | -682.963 | -682.860(b2) | -682.195 |
| Mar | Ottawa CDA | -612.873 | -595.826 | -595.120 | -597.433 | -592.844(b2) | -594.627 |
| Mar | St. Alban | -584.381 | -576.554 | -576.553 | -578.947 | -575.217(b2) | -575.092 |
| Mar | St. Jerome | -630.438 | -621.225 | -620.765 | -619.149 | -618.994(b2) | -620.593 |
| Apr | Dorval | -691.784 | -690.566 | -691.264 | -691.784(a) | CV<1 | -688.451 |
| Apr | Cornwall | -676.497 | -676.463 | -676.373 | -676.497(a) | CV<1 | -675.588 |
| Apr | Drummondville | -740.257 | -740.203 | -740.234 | -740.257(a) | CV<1 | -737.578 |
| Apr | Farnham | -808.007 | -806.317 | -807.021 | -807.960 | -802.884(b3) | -803.830 |
| Apr | Lennoxville | -761.923 | -760.546 | -760.690 | -759.702 | -760.928(b3) | -757.577 |
| Apr | Morrisburg | -730.451 | -727.373 | -728.163 | -722.844 | -724.655(b1) | -726.683 |
| Apr | Oka | -708.898 | -708.896 | -708.881 | -708.898(a) | CV<1 | -705.782 |
| Apr | Ottawa CDA | -647.523 | -645.383 | -646.098 | -642.751 | -646.373(b2) | -644.388 |
| Apr | St. Alban | -642.800 | -641.335 | -640.897 | -640.607 | -640.501(b1) | -640.052 |
| Apr | St. Jerome | -647.927 | -643.726 | -644.059 | -643.039 | -643.129(b2) | -643.007 |
| May | Dorval | -690.347 | -684.055 | -684.173 | -682.447 | -682.729(b3) | -682.067 |
| May | Cornwall | -694.898 | -690.837 | -691.946 | -686.347 | -691.074(b2) | -689.497 |
| May | Drummondville | -748.045 | -745.700 | -745.661 | -745.678 | -743.667(b2) | -743.379 |
| May | Farnham | -805.953 | -805.203 | -805.685 | -805.953(a) | CV<1 | -803.135 |
| May | Lennoxville | -841.174 | -839.438 | -839.966 | -839.344 | -840.936(b1) | -838.218 |
| May | Morrisburg | -732.892 | -730.542 | -730.624 | -730.045 | -730.014(b2) | -727.661 |
| May | Oka | -727.454 | -726.561 | -726.663 | -726.471 | -726.416(b1) | -726.346 |
| May | Ottawa CDA | -705.469 | -703.495 | -704.098 | -699.252 | -702.534(b2) | -702.803 |
| May | St. Alban | -840.218 | -839.353 | -839.010 | -838.356 | -838.383(b3) | -837.436 |
| May | St. Jerome | -701.800 | -698.832 | -699.397 | -696.140 | -696.297(b2) | -698.567 |
| Jun | Dorval | -712.453 | -707.826 | -707.476 | -707.433 | -706.929(b2) | -707.241 |
| Jun | Cornwall | -664.474 | -657.468 | -658.079 | -659.390 | -653.879(b2) | -656.384 |
| Jun | Drummondville | -763.526 | -758.574 | -759.053 | -756.437 | -757.752(b2) | -757.334 |
| Jun | Farnham | -806.823 | -795.700 | -797.081 | -791.046 | -790.469(b2) | -794.107 |
| Jun | Lennoxville | -835.827 | -833.440 | -833.587 | -833.457 | -833.068(b2) | -832.871 |
| Jun | Morrisburg | -685.465 | -679.850 | -679.587 | -680.187 | -679.343(b2) | -678.812 |
| Jun | Oka | -755.563 | -754.936 | -754.503 | -753.818 | -753.822(b3) | -752.819 |
| Jun | Ottawa CDA | -738.634 | -725.375 | -725.402 | -726.207 | -724.288(b2) | -724.683 |
| Jun | St. Alban | -867.105 | -858.757 | -858.312 | -858.192 | -857.769(b2) | -855.845 |
| Jun | St. Jerome | -674.684 | -659.887 | -659.706 | -659.465 | -657.727(b2) | -657.272 |
| Jul | Dorval | -757.307 | -746.819 | -747.255 | -746.873 | -746.118(b2) | -745.648 |
| Jul | Cornwall | -677.956 | -669.800 | -670.718 | -667.194 | -667.024(b2) | -668.845 |
| Jul | Drummondville | -811.623 | -805.273 | -804.809 | -806.040 | -804.590(b1) | -804.739 |
| Jul | Farnham | -915.703 | -906.798 | -905.612 | -906.207 | -905.323(b2) | -905.339 |
| Jul | Lennoxville | -936.304 | -930.094 | -930.417 | -931.396 | -928.856(b2) | -929.666 |
| Jul | Morrisburg | -721.638 | -718.404 | -718.448 | -718.307 | -717.787(b2) | -717.343 |
| Jul | Oka | -746.707 | -742.356 | -742.844 | -742.577 | -742.338 | -740.875 |
| Jul | Ottawa CDA | -752.419 | -741.844 | -742.914 | -742.268 | -741.302(b2) | -740.028 |
| Jul | St. Alban | -872.552 | -863.714 | -864.910 | -863.304 | -862.876 | -862.914 |
| Jul | St. Jerome | -740.128 | -729.541 | -730.323 | -728.887 | -728.715 | -728.644 |
| Aug | Dorval | -788.834 | -778.709 | -779.232 | -780.242 | -777.478(b2) | -778.583 |
| Aug | Cornwall | -748.197 | -732.445 | -733.150 | -733.688 | -731.027(b2) | -732.250 |
| Aug | Drummondville | -824.701 | -813.022 | -811.456 | -811.403 | -810.701(b2) | -808.286 |
| Aug | Farnham | -953.282 | -938.256 | -936.646 | -936.628 | -936.026 | -935.289 |
| Aug | Lennoxville | -918.167 | -901.978 | -902.724 | -899.001 | -897.048(b2) | -901.354 |
| Aug | Morrisburg | -737.797 | -727.823 | -729.121 | -726.984 | -726.275(b2) | -726.265 |
| Aug | Oka | -850.929 | -834.250 | -834.379 | -835.561 | -832.096(b2) | -832.798 |
| Aug | Ottawa CDA | -717.012 | -701.209 | -701.201 | -700.726 | -699.976 | -699.738 |
| Aug | St. Alban | -922.172 | -906.862 | -902.956 | -901.609 | -901.218(b2) | -900.801 |
| Aug | St. Jerome | -807.337 | -790.934 | -790.576 | -789.405 | -789.027(b1) | -788.026 |
| Sep | Dorval | -694.147 | -685.409 | -684.043 | -684.845 | -683.048(b1) | -682.228 |
| Sep | Cornwall | -703.378 | -688.702 | -687.550 | -691.815 | -686.419(b1) | -687.089 |
| Sep | Drummondville | -748.328 | -734.492 | -733.317 | -734.289 | -733.020(b2) | -733.010 |
| Sep | Farnham | -765.469 | -750.719 | -752.866 | -745.786 | -745.079(b1) | -745.627 |
| Sep | Lennoxville | -756.144 | -745.815 | -746.752 | -742.889 | -742.732(b1) | -745.700 |
| Sep | Morrisburg | -703.659 | -690.030 | -687.744 | -688.516 | -686.797(b2) | -686.818 |
| Sep | Oka | -727.353 | -716.206 | -715.957 | -715.125 | -715.231 | -714.846 |
| Sep | Ottawa CDA | -730.038 | -705.716 | -706.545 | -705.728 | -702.882(b2) | -704.080 |
| Sep | St. Alban | -849.683 | -841.584 | -839.718 | -835.916 | -835.934(b3) | -834.735 |
| Sep | St. Jerome | -733.822 | -718.605 | -719.559 | -719.066 | -717.364(b2) | -717.952 |
| Oct | Dorval | -689.902 | -685.767 | -686.745 | -682.543 | -683.426(b3) | -685.119 |
| Oct | Cornwall | -644.790 | -641.529 | -641.024 | -641.143 | -640.624(b1) | -639.766 |
| Oct | Drummondville | -749.074 | -743.762 | -744.911 | -739.241 | -739.207(b2) | -742.934 |
| Oct | Farnham | -812.975 | -809.629 | -809.052 | -808.284 | -806.975(b1) | -807.294 |
| Oct | Lennoxville | -764.267 | -763.486 | -763.114 | -761.955 | -761.995(b3) | -759.204 |
| Oct | Morrisburg | -684.531 | -676.008 | -675.930 | -678.889 | -674.508(b2) | -673.543 |
| Oct | Oka | -698.792 | -693.929 | -693.385 | -693.989 | -692.357(b2) | -691.258 |
| Oct | Ottawa CDA | -702.224 | -694.792 | -694.881 | -694.757 | -694.176(b2) | -694.359 |
| Oct | St. Alban | -787.214 | -785.100 | -785.150 | -783.862 | -784.016(b2) | -785.082 |
| Oct | St. Jerome | -751.596 | -749.455 | -749.187 | -748.114 | -747.473(b2) | -745.549 |
| Nov | Dorval | -833.409 | -828.663 | -829.670 | -829.533 | -827.117(b2) | -825.831 |
| Nov | Cornwall | -795.342 | -794.056 | -793.999 | -794.142 | -793.976(b1) | -793.353 |
| Nov | Drummondville | -933.792 | -926.925 | -928.889 | -918.485 | -920.949(b2) | -924.891 |
| Nov | Farnham | -932.457 | -927.647 | -929.666 | -932.457(a) | CV<1 | -923.715 |
| Nov | Lennoxville | -875.282 | -865.951 | -865.921 | -866.436 | -864.435(b2) | -864.202 |
| Nov | Morrisburg | -828.655 | -826.244 | -826.819 | -827.675 | -820.271(b2) | -822.826 |
| Nov | Oka | -855.354 | -853.789 | -853.971 | -854.177 | -852.675(b2) | -851.813 |
| Nov | Ottawa CDA | -790.376 | -780.586 | -782.105 | -775.155 | -775.034(b2) | -779.139 |
| Nov | St. Alban | -773.407 | -773.235 | -773.284 | -773.407(a) | -773.391(b1) | -772.396 |
| Nov | St. Jerome | -830.801 | -826.628 | -827.305 | -825.427 | -825.940(b3) | -825.120 |
| Dec | Dorval | -863.632 | -849.641 | -849.315 | -849.216 | -848.598 | -848.592 |
| Dec | Cornwall | -854.923 | -849.735 | -850.149 | -850.458 | -849.657 | -849.017 |
| Dec | Drummondville | -932.154 | -923.483 | -923.961 | -922.938 | -921.528(b2) | -920.812 |
| Dec | Farnham | -996.702 | -989.012 | -989.340 | -992.280 | -983.908(b2) | -987.835 |
| Dec | Lennoxville | -935.718 | -922.100 | -922.026 | -919.060 | -917.691(b2) | -920.399 |
| Dec | Morrisburg | -914.699 | -903.381 | -904.674 | -898.592 | -897.922(b2) | -902.793 |
| Dec | Oka | -917.793 | -917.088 | -917.028 | -916.785 | -916.791(b3) | -916.660 |
| Dec | Ottawa CDA | -787.470 | -779.181 | -779.335 | -777.942 | -778.122(b3) | -775.266 |
| Dec | St. Alban | -886.594 | -883.495 | -882.435 | -881.023 | -881.030(b3) | -882.435(c) |
| Dec | St. Jerome | -875.703 | -872.353 | -872.147 | -873.035 | -871.996(b2) | -871.051 |

Table 1. The values of the logarithm of the ML of all the candidate distributions. (a) The corresponding ML estimation of the mixed exponential distribution results in the ML estimation of the exponential distribution. (b1) The corresponding mixture estimation of the MGW distribution results in the mixture estimation of the MGE distribution. (b2) The corresponding mixture estimation of the MGW distribution results in the mixture estimation of the MEW distribution. (b3) The corresponding mixture estimation of the MGW distribution results in the mixture estimation of the mixed exponential distribution. (c) The corresponding ML estimation of the MGW distribution results in the ML estimation of the Weibull distribution.

Based on Table 1, likelihood ratio tests can be performed by taking the values of the logarithm of the ML of the MGW (ML estimation) distribution as the comparative base. For the ML estimation of the exponential, Gamma, and Weibull distributions, the likelihood ratio test statistic follows the chi-squared distribution with four, three, and three degrees of freedom respectively. For the ML estimation of the mixed exponential distribution, the likelihood ratio test statistic follows the chi-squared distribution with two degrees of freedom, and the likelihood ratio test statistic follows the chi-squared distribution with four degrees of freedom when the ML estimation of the mixed exponential distribution results in the ML estimation of the exponential distribution. For the mixture estimation of the MGW distribution, the likelihood ratio test statistic follows the chi-squared distribution with two degrees of freedom because two constraints have been added before the ML estimation step of the mixture estimation, and only three free parameters are adopted for the ML estimation part of the mixture estimation. Similarly, when the mixture estimation of the MGW distribution results in the mixture estimation of the MEW, MGE, and mixed exponential distribution respectively, the chi-squared distribution with three, three, and four degrees of freedom should be adopted. However, the likelihood ratio test cannot be applied (except for the exponential distribution with the ML estimation case) when the ML estimation of the MGW distribution results in the ML estimation of the Gamma/Weibull distribution. Similarly, the likelihood ratio test cannot be applied when the value of the ML of the candidate distribution is larger than the value of the ML of the MGW (ML estimation) distribution. Finally, based on the above, the p-values of the results of the likelihood ratio test can be summarized in Table 2.

| Month | Site | Exponential | Gamma | Weibull | Mixed Exponential | MGW (mixture estimation) |
|---|---|---|---|---|---|---|
| Jan | Dorval | 0.0000 | 0.0134 | 0.1099 | 0.0774 | 0.3828 |
| Jan | Cornwall | 0.0008 | 0.0119 | 0.0455 | 0.2843 | 0.5290 |
| Jan | Drummondville | 0.0028 | not applicable | not applicable | not applicable | not applicable |
| Jan | Farnham | 0.0000 | 0.7844 | 0.1903 | not applicable | not applicable |
| Jan | Lennoxville | 0.0000 | 0.0116 | 0.0794 | 0.6387 | 0.9147 |
| Jan | Morrisburg | 0.0000 | 0.0825 | 0.6703 | 0.1117 | not applicable |
| Jan | Oka | 0.0015 | 0.0981 | 0.2332 | 0.1098 | 0.5421 |
| Jan | Ottawa CDA | 0.0000 | 0.0677 | 0.2254 | 0.0188 | 0.4366 |
| Jan | St. Alban | 0.0675 | 0.0809 | 0.1542 | 0.4135 | 0.7693 |
| Jan | St. Jerome | 0.0000 | 0.0010 | 0.0077 | 0.2035 | 0.4853 |
| Feb | Dorval | 0.0000 | 0.0469 | 0.0639 | 0.0373 | 0.1182 |
| Feb | Cornwall | 0.0668 | 0.6008 | 0.4987 | 0.9462 | 0.9821 |
| Feb | Drummondville | 0.0004 | 0.0555 | 0.1345 | 0.0288 | 0.5088 |
| Feb | Farnham | 0.0000 | 0.0705 | 0.0063 | not applicable | not applicable |
| Feb | Lennoxville | 0.0000 | 0.2089 | 0.1073 | 0.8461 | 0.9904 |
| Feb | Morrisburg | 0.0429 | 0.3449 | 0.5614 | 0.4957 | 0.8124 |
| Feb | Oka | 0.5456 | 0.6456 | 0.6125 | 0.5361 | 0.6191 |
| Feb | Ottawa CDA | 0.0000 | 0.0124 | 0.0745 | 0.0292 | 0.2967 |
| Feb | St. Alban | 0.0003 | 0.3603 | 0.4231 | 0.9742 | 0.9813 |
| Feb | St. Jerome | 0.0132 | 0.1672 | 0.3333 | 0.5679 | 0.8733 |
| Mar | Dorval | 0.0001 | not applicable | not applicable | not applicable | not applicable |
| Mar | Cornwall | 0.0075 | 0.8238 | 0.8242 | 0.3127 | not applicable |
| Mar | Drummondville | 0.0001 | 0.0293 | 0.0481 | 0.0188 | 0.0722 |
| Mar | Farnham | 0.0332 | 0.4921 | 0.2027 | not applicable | 0.6179 |
| Mar | Lennoxville | 0.0258 | 0.6883 | 0.6340 | not applicable | not applicable |
| Mar | Morrisburg | 0.1320 | 0.4923 | 0.2792 | not applicable | 0.2365 |
| Mar | Oka | 0.0377 | 0.6617 | 0.4790 | 0.4644 | 0.7223 |
| Mar | Ottawa CDA | 0.0000 | 0.4938 | 0.8048 | 0.0604 | not applicable |
| Mar | St. Alban | 0.0010 | 0.4034 | 0.4039 | 0.0212 | 0.9690 |
| Mar | St. Jerome | 0.0006 | 0.7377 | 0.9514 | not applicable | not applicable |
| Apr | Dorval | 0.1546 | 0.2376 | 0.1312 | 0.1546 | CV<1 |
| Apr | Cornwall | 0.7692 | 0.6257 | 0.6661 | 0.7692 | CV<1 |
| Apr | Drummondville | 0.2524 | 0.1543 | 0.1503 | 0.2524 | CV<1 |
| Apr | Farnham | 0.0795 | 0.1737 | 0.0944 | 0.0161 | not applicable |
| Apr | Lennoxville | 0.0693 | 0.1146 | 0.1011 | 0.1194 | 0.1524 |
| Apr | Morrisburg | 0.1101 | 0.7101 | 0.3978 | not applicable | not applicable |
| Apr | Oka | 0.1825 | 0.1011 | 0.1024 | 0.1825 | CV<1 |
| Apr | Ottawa CDA | 0.1799 | 0.5746 | 0.3313 | not applicable | 0.2646 |
| Apr | St. Alban | 0.2401 | 0.4636 | 0.6392 | 0.5740 | 0.8263 |
| Apr | St. Jerome | 0.0432 | 0.6963 | 0.5510 | 0.9685 | 0.9701 |
| May | Dorval | 0.0024 | 0.2641 | 0.2395 | 0.6842 | 0.8574 |
| May | Cornwall | 0.0289 | 0.4436 | 0.1794 | not applicable | 0.3685 |
| May | Drummondville | 0.0533 | 0.2000 | 0.2066 | 0.1004 | 0.9019 |
| May | Farnham | 0.2279 | 0.2470 | 0.1646 | 0.2279 | CV<1 |
| May | Lennoxville | 0.2059 | 0.4862 | 0.3215 | 0.3245 | 0.1425 |
| May | Morrisburg | 0.0333 | 0.1238 | 0.1153 | 0.0922 | 0.1946 |
| May | Oka | 0.6958 | 0.9337 | 0.8883 | 0.8823 | 0.9864 |
| May | Ottawa CDA | 0.2549 | 0.7092 | 0.4594 | not applicable | not applicable |
| May | St. Alban | 0.2341 | 0.2800 | 0.3694 | 0.3986 | 0.7555 |
| May | St. Jerome | 0.1669 | 0.9121 | 0.6458 | not applicable | not applicable |
| Jun | Dorval | 0.0339 | 0.7600 | 0.9253 | 0.8249 | not applicable |
| Jun | Cornwall | 0.0028 | 0.5380 | 0.3351 | 0.0495 | not applicable |
| Jun | Drummondville | 0.0147 | 0.4791 | 0.3289 | not applicable | 0.8410 |
| Jun | Farnham | 0.0000 | 0.3639 | 0.1142 | not applicable | not applicable |
| Jun | Lennoxville | 0.2059 | 0.7678 | 0.6980 | 0.5566 | 0.9414 |
| Jun | Morrisburg | 0.0099 | 0.5569 | 0.6711 | 0.2529 | 0.7865 |
| Jun | Oka | 0.2408 | 0.2373 | 0.3385 | 0.3683 | 0.7349 |
| Jun | Ottawa CDA | 0.0000 | 0.7094 | 0.6970 | 0.2179 | not applicable |
| Jun | St. Alban | 0.0002 | 0.1205 | 0.1767 | 0.0957 | 0.2784 |
| Jun | St. Jerome | 0.0000 | 0.1557 | 0.1817 | 0.1115 | 0.8227 |
| Jul | Dorval | 0.0001 | 0.5046 | 0.3598 | 0.2938 | 0.8155 |
| Jul | Cornwall | 0.0011 | 0.5914 | 0.2903 | not applicable | not applicable |
| Jul | Drummondville | 0.0081 | 0.7848 | 0.9867 | 0.2723 | not applicable |
| Jul | Farnham | 0.0004 | 0.4046 | 0.9088 | 0.4199 | not applicable |
| Jul | Lennoxville | 0.0100 | 0.8361 | 0.6817 | 0.1772 | not applicable |
| Jul | Morrisburg | 0.0722 | 0.5476 | 0.5303 | 0.3815 | 0.8288 |
| Jul | Oka | 0.0200 | 0.3975 | 0.2683 | 0.1823 | 0.2316 |
| Jul | Ottawa CDA | 0.0001 | 0.3039 | 0.1232 | 0.1065 | 0.4667 |
| Jul | St. Alban | 0.0007 | 0.6594 | 0.2623 | 0.6767 | not applicable |
| Jul | St. Jerome | 0.0001 | 0.6159 | 0.3397 | 0.7840 | 0.9313 |
| Aug | Dorval | 0.0004 | 0.9687 | 0.7296 | 0.1903 | not applicable |
| Aug | Cornwall | 0.0000 | 0.9424 | 0.6151 | 0.2374 | not applicable |
| Aug | Drummondville | 0.0000 | 0.0236 | 0.0961 | 0.0443 | 0.1846 |
| Aug | Farnham | 0.0000 | 0.1148 | 0.4377 | 0.2620 | 0.4787 |
| Aug | Lennoxville | 0.0000 | 0.7419 | 0.4337 | not applicable | not applicable |
| Aug | Morrisburg | 0.0001 | 0.3741 | 0.1265 | 0.4871 | 0.9993 |
| Aug | Oka | 0.0000 | 0.4065 | 0.3671 | 0.0631 | not applicable |
| Aug | Ottawa CDA | 0.0000 | 0.4006 | 0.4031 | 0.3726 | 0.7880 |
| Aug | St. Alban | 0.0000 | 0.0070 | 0.2298 | 0.4459 | 0.8412 |
| Aug | St. Jerome | 0.0000 | 0.1208 | 0.1645 | 0.2518 | 0.5719 |
| Sep | Dorval | 0.0001 | 0.0953 | 0.3042 | 0.0730 | 0.6501 |
| Sep | Cornwall | 0.0000 | 0.3581 | 0.8204 | 0.0089 | not applicable |
| Sep | Drummondville | 0.0000 | 0.3973 | 0.8932 | 0.2784 | 0.9993 |
| Sep | Farnham | 0.0000 | 0.0171 | 0.0023 | 0.8529 | not applicable |
| Sep | Lennoxville | 0.0003 | 0.9725 | 0.5509 | not applicable | not applicable |
| Sep | Morrisburg | 0.0000 | 0.0927 | 0.6035 | 0.1830 | not applicable |
| Sep | Oka | 0.0000 | 0.4368 | 0.5276 | 0.7568 | 0.6804 |
| Sep | Ottawa CDA | 0.0000 | 0.3515 | 0.1769 | 0.1923 | not applicable |
| Sep | St. Alban | 0.0000 | 0.0033 | 0.0189 | 0.3070 | 0.6630 |
| Sep | St. Jerome | 0.0000 | 0.7274 | 0.3598 | 0.3283 | not applicable |
| Oct | Dorval | 0.0484 | 0.7302 | 0.3544 | not applicable | not applicable |
| Oct | Cornwall | 0.0396 | 0.3174 | 0.4726 | 0.2523 | 0.6334 |
| Oct | Drummondville | 0.0154 | 0.6464 | 0.2664 | not applicable | not applicable |
| Oct | Farnham | 0.0228 | 0.1976 | 0.3185 | 0.3716 | not applicable |
| Oct | Lennoxville | 0.0384 | 0.0357 | 0.0499 | 0.0639 | 0.2326 |
| Oct | Morrisburg | 0.0002 | 0.1771 | 0.1892 | 0.0048 | 0.5870 |
| Oct | Oka | 0.0046 | 0.1484 | 0.2353 | 0.0652 | 0.5321 |
| Oct | Ottawa CDA | 0.0034 | 0.8338 | 0.7904 | 0.6715 | not applicable |
| Oct | St. Alban | 0.3715 | 0.9981 | 0.9871 | not applicable | not applicable |
| Oct | St. Jerome | 0.0167 | 0.0500 | 0.0636 | 0.0769 | 0.2782 |
| Nov | Dorval | 0.0044 | 0.1291 | 0.0532 | 0.0247 | 0.4624 |
| Nov | Cornwall | 0.4090 | 0.7044 | 0.7311 | 0.4543 | 0.7422 |
| Nov | Drummondville | 0.0013 | 0.2540 | 0.0461 | not applicable | not applicable |
| Nov | Farnham | 0.0016 | 0.0489 | 0.0077 | 0.0016 | CV<1 |
| Nov | Lennoxville | 0.0002 | 0.3210 | 0.3288 | 0.1071 | 0.9263 |
| Nov | Morrisburg | 0.0201 | 0.0773 | 0.0463 | 0.0078 | not applicable |
| Nov | Oka | 0.1316 | 0.2668 | 0.2294 | 0.0941 | 0.6317 |
| Nov | Ottawa CDA | 0.0002 | 0.4082 | 0.1149 | not applicable | not applicable |
| Nov | St. Alban | 0.7319 | 0.6422 | 0.6205 | 0.7319 | 0.5747 |
| Nov | St. Jerome | 0.0228 | 0.3891 | 0.2242 | 0.7355 | 0.8017 |
| Dec | Dorval | 0.0000 | 0.5524 | 0.6949 | 0.5358 | 0.9949 |
| Dec | Cornwall | 0.0188 | 0.6973 | 0.5195 | 0.2368 | 0.5276 |
| Dec | Drummondville | 0.0001 | 0.1484 | 0.0980 | 0.1193 | 0.6981 |
| Dec | Farnham | 0.0014 | 0.5023 | 0.3901 | 0.0117 | not applicable |
| Dec | Lennoxville | 0.0000 | 0.3336 | 0.3542 | not applicable | not applicable |
| Dec | Morrisburg | 0.0001 | 0.7587 | 0.2883 | not applicable | not applicable |
| Dec | Oka | 0.6870 | 0.8362 | 0.8648 | 0.8828 | 0.9922 |
| Dec | Ottawa CDA | 0.0001 | 0.0497 | 0.0433 | 0.0688 | 0.2216 |
| Dec | St. Alban | 0.0039 | not applicable | not applicable | not applicable | not applicable |
| Dec | St. Jerome | 0.0539 | 0.4567 | 0.5333 | 0.1375 | 0.5952 |

Table 2. The p-values of the results of the likelihood ratio test.

However, the inability to apply the likelihood ratio test does not necessarily indicate the failure of the corresponding candidate distributions. On the contrary, considering the physically reasonable shapes of the PDFs of the candidate distributions and their superior fit to the ML estimation result of the MGW distribution, the corresponding candidate distributions should be selected as the final model when the likelihood ratio test cannot be performed. The reason that a candidate distribution has a larger ML value than the corresponding ML estimation result of the MGW distribution is because the estimated parameter of the candidate distribution is located at the boundary of our defined five-dimensional MGW parameter space, and the nearest local optimum to the estimated parameter of the candidate distribution is located outside of the defined parameter space. As a result, it is reasonable to view the estimated parameters of the candidate distributions that have larger ML values than the corresponding ML estimation result of the MGW distribution as consistent estimators of the true parameter; and the MGW distribution framework should be replaced by a new family of distributions to perform additional likelihood ratio tests. Perhaps, a family of distributions that only has monotonically decreasing PDFs with a four-dimensional parameter space would be ideal. However, this new family of distributions is still difficult to find in practice. Nonetheless, the Akaike (1972) information criterion (AIC) provides an alternative solution by looking at the second-order variation of the logarithm of likelihood ratios (or so-called information) when the estimated parameters are deemed to be consistent estimators of the true parameter. Finally, by using the AIC as a supplementary of the likelihood ratio test, the model selection for the daily precipitation amounts under the MGW distribution framework can be done, and the result is summarized in Table 3.

| Month | Site | Selected Model | $p$ | $\alpha$ | $\beta$ | $k$ | $\lambda$ |
|---|---|---|---|---|---|---|---|
| Jan | Dorval | MGE (mixture estimation) | 0.4600 | 0.8190 | 1.9057 | 1.0000 | 7.5273 |
| Jan | Cornwall | MGE (mixture estimation) | 0.8600 | 0.9426 | 4.1319 | 1.0000 | 12.0856 |
| Jan | Drummondville | MEW (mixture estimation) | 0.1900 | 1.0000 | 1.4017 | 0.9560 | 5.8720 |
| Jan | Farnham | MEW (mixture estimation) | 0.1000 | 1.0000 | 0.0625 | 0.9711 | 4.7918 |
| Jan | Lennoxville | Mixed Exponential (mixture estimation) | 0.6000 | 1.0000 | 6.4231 | 1.0000 | 1.5545 |
| Jan | Morrisburg | MGW (mixture estimation) | 0.0400 | 0.9157 | 17.9579 | 0.8313 | 3.6968 |
| Jan | Oka | MGE (mixture estimation) | 0.8300 | 0.8653 | 5.1800 | 1.0000 | 12.2031 |
| Jan | Ottawa CDA | MGE (mixture estimation) | 0.7600 | 0.7901 | 3.2340 | 1.0000 | 7.5287 |
| Jan | St. Alban | Mixed Exponential (mixture estimation) | 0.5700 | 1.0000 | 3.4343 | 1.0000 | 8.2872 |
| Jan | St. Jerome | Mixed Exponential (mixture estimation) | 0.4200 | 1.0000 | 2.3481 | 1.0000 | 9.4245 |
| Feb | Dorval | MEW (mixture estimation) | 0.7500 | 1.0000 | 6.6020 | 0.8632 | 1.0380 |
| Feb | Cornwall | MGE (mixture estimation) | 0.9000 | 0.9901 | 6.1019 | 1.0000 | 0.5701 |
| Feb | Drummondville | MGE (mixture estimation) | 0.9600 | 0.8116 | 5.7966 | 1.0000 | 20.0121 |
| Feb | Farnham | MEW (mixture estimation) | 0.0900 | 1.0000 | 0.0590 | 0.9415 | 5.7528 |
| Feb | Lennoxville | MEW (mixture estimation) | 0.1400 | 1.0000 | 0.2977 | 0.9530 | 5.8644 |
| Feb | Morrisburg | MGE (mixture estimation) | 0.5800 | 0.9070 | 3.6714 | 1.0000 | 7.8348 |
| Feb | Oka | Gamma (ML estimation) | | 0.9078 | 6.8094 | | |
| Feb | Ottawa CDA | MEW (mixture estimation) | 0.0400 | 1.0000 | 0.0552 | 0.7734 | 3.9314 |
| Feb | St. Alban | MGW (mixture estimation) | 0.1800 | 0.9901 | 0.7146 | 0.9620 | 6.5826 |
| Feb | St. Jerome | Mixed Exponential (mixture estimation) | 0.2800 | 1.0000 | 2.1166 | 1.0000 | 8.8028 |
| Mar | Dorval | MGE (mixture estimation) | 0.3000 | 0.8653 | 2.4134 | 1.0000 | 8.3565 |
| Mar | Cornwall | MEW (mixture estimation) | 0.0500 | 1.0000 | 0.0794 | 0.8962 | 5.8661 |
| Mar | Drummondville | MEW (mixture estimation) | 0.0300 | 1.0000 | 0.1051 | 0.8654 | 6.3150 |
| Mar | Farnham | Mixed Exponential (ML estimation) | 0.0464 | 1.0000 | 0.0555 | 1.0000 | 6.3447 |
| Mar | Lennoxville | MGE (mixture estimation) | 0.8800 | 0.9901 | 5.9756 | 1.0000 | 0.6698 |
| Mar | Morrisburg | Mixed Exponential (ML estimation) | 0.0392 | 1.0000 | 0.0537 | 1.0000 | 6.1612 |
| Mar | Oka | MEW (mixture estimation) | 0.9100 | 1.0000 | 8.1688 | 0.8723 | 0.6477 |
| Mar | Ottawa CDA | MEW (mixture estimation) | 0.0500 | 1.0000 | 0.0581 | 0.7920 | 4.8707 |
| Mar | St. Alban | MEW (mixture estimation) | 0.0400 | 1.0000 | 0.1057 | 0.8609 | 6.2236 |
| Mar | St. Jerome | MEW (mixture estimation) | 0.1300 | 1.0000 | 0.5305 | 0.9115 | 7.0297 |
| Apr | Dorval | Gamma (ML estimation) | | 0.8836 | 8.1077 | | |
| Apr | Cornwall | Exponential (ML estimation) | | 1.0000 | 6.6261 | 1.0000 | 6.6261 |
| Apr | Drummondville | Exponential (ML estimation) | | 1.0000 | 6.4118 | 1.0000 | 6.4118 |
| Apr | Farnham | Mixed Exponential (mixture estimation) | 0.9800 | 1.0000 | 6.8650 | 1.0000 | 0.0882 |
| Apr | Lennoxville | Mixed Exponential (mixture estimation) | 0.0300 | 1.0000 | 1.1787 | 1.0000 | 6.6150 |
| Apr | Morrisburg | Mixed Exponential (ML estimation) | 0.0444 | 1.0000 | 0.0514 | 1.0000 | 6.8268 |
| Apr | Oka | Exponential (ML estimation) | | 1.0000 | 7.2324 | 1.0000 | 7.2324 |
| Apr | Ottawa CDA | Mixed Exponential (ML estimation) | 0.0383 | 1.0000 | 0.0615 | 1.0000 | 6.3078 |
| Apr | St. Alban | MGE (mixture estimation) | 0.7000 | 0.9426 | 5.2789 | 1.0000 | 10.5759 |
| Apr | St. Jerome | MEW (mixture estimation) | 0.0200 | 1.0000 | 0.0609 | 0.8938 | 7.8891 |
| May | Dorval | Mixed Exponential (mixture estimation) | 0.8700 | 1.0000 | 6.3725 | 1.0000 | 0.6844 |
| May | Cornwall | Mixed Exponential (ML estimation) | 0.0499 | 1.0000 | 0.0533 | 1.0000 | 6.4502 |
| May | Drummondville | MEW (mixture estimation) | 0.0500 | 1.0000 | 0.0640 | 0.9590 | 6.2574 |
| May | Farnham | Gamma (ML estimation) | | 0.9151 | 6.8657 | | |
| May | Lennoxville | Gamma (ML estimation) | | 0.8750 | 8.1286 | | |
| May | Morrisburg | MEW (mixture estimation) | 0.0200 | 1.0000 | 0.0424 | 0.9650 | 5.9323 |
| May | Oka | MGE (mixture estimation) | 0.9800 | 0.9518 | 6.2665 | 1.0000 | 0.1060 |
| May | Ottawa CDA | Mixed Exponential (ML estimation) | 0.0419 | 1.0000 | 0.0611 | 1.0000 | 6.8340 |
| May | St. Alban | Mixed Exponential (mixture estimation) | 0.3000 | 1.0000 | 4.0483 | 1.0000 | 9.6740 |
| May | St. Jerome | MEW (mixture estimation) | 0.0300 | 1.0000 | 0.0681 | 0.9711 | 7.8102 |
| Jun | Dorval | MEW (mixture estimation) | 0.0200 | 1.0000 | 0.0791 | 0.8793 | 7.0253 |
| Jun | Cornwall | MEW (mixture estimation) | 0.0600 | 1.0000 | 0.0575 | 0.9248 | 7.2292 |
| Jun | Drummondville | Mixed Exponential (ML estimation) | 0.0413 | 1.0000 | 0.0549 | 1.0000 | 7.3145 |
| Jun | Farnham | MEW (mixture estimation) | 0.0600 | 1.0000 | 0.0494 | 0.9038 | 7.4800 |
| Jun | Lennoxville | MEW (mixture estimation) | 0.0200 | 1.0000 | 0.0532 | 0.9530 | 7.4263 |
| Jun | Morrisburg | MEW (mixture estimation) | 0.0400 | 1.0000 | 0.0681 | 0.8654 | 6.3421 |
| Jun | Oka | Mixed Exponential (mixture estimation) | 0.6300 | 1.0000 | 4.9856 | 1.0000 | 10.1041 |
| Jun | Ottawa CDA | MEW (mixture estimation) | 0.0400 | 1.0000 | 0.1147 | 0.8437 | 7.0552 |
| Jun | St. Alban | MEW (mixture estimation) | 0.0400 | 1.0000 | 0.1075 | 0.8840 | 7.5524 |
| Jun | St. Jerome | MEW (mixture estimation) | 0.0500 | 1.0000 | 0.0612 | 0.7718 | 6.5839 |
| Jul | Dorval | MEW (mixture estimation) | 0.0400 | 1.0000 | 0.0887 | 0.9013 | 7.4439 |
| Jul | Cornwall | MEW (mixture estimation) | 0.0700 | 1.0000 | 0.0664 | 0.9168 | 7.6198 |
| Jul | Drummondville | MGE (mixture estimation) | 0.8500 | 0.8116 | 7.7711 | 1.0000 | 14.9410 |
| Jul | Farnham | MEW (mixture estimation) | 0.1600 | 1.0000 | 4.0095 | 0.8313 | 8.9528 |
| Jul | Lennoxville | MEW (mixture estimation) | 0.0300 | 1.0000 | 0.1195 | 0.8889 | 8.2831 |
| Jul | Morrisburg | MEW (mixture estimation) | 0.0200 | 1.0000 | 0.1543 | 0.9248 | 8.3389 |
| Jul | Oka | Gamma (ML estimation) | | 0.8012 | 9.3236 | | |
| Jul | Ottawa CDA | MEW (mixture estimation) | 0.0600 | 1.0000 | 0.0449 | 0.9501 | 8.0420 |
| Jul | St. Alban | MGW (mixture estimation) | 0.9400 | 0.9246 | 11.2334 | 0.9358 | 0.2685 |
| Jul | St. Jerome | MGW (mixture estimation) | 0.9100 | 0.8985 | 9.0763 | 0.9773 | 0.3584 |
| Aug | Dorval | MEW (mixture estimation) | 0.0300 | 1.0000 | 0.0382 | 0.8816 | 8.5737 |
| Aug | Cornwall | MEW (mixture estimation) | 0.0600 | 1.0000 | 0.1284 | 0.8632 | 9.3861 |
| Aug | Drummondville | MEW (mixture estimation) | 0.5300 | 1.0000 | 12.9391 | 0.8677 | 3.3816 |
| Aug | Farnham | MGW (mixture estimation) | 0.1700 | 0.9426 | 1.9164 | 0.8565 | 9.2733 |
| Aug | Lennoxville | MEW (mixture estimation) | 0.0500 | 1.0000 | 0.0380 | 0.8632 | 8.6129 |
| Aug | Morrisburg | MEW (mixture estimation) | 0.0400 | 1.0000 | 0.0731 | 0.9089 | 8.9252 |
| Aug | Oka | MEW (mixture estimation) | 0.0500 | 1.0000 | 0.0492 | 0.8565 | 8.3564 |
| Aug | Ottawa CDA | MGW (mixture estimation) | 0.7300 | 0.9901 | 10.7867 | 0.8214 | 1.3397 |
| Aug | St. Alban | MEW (mixture estimation) | 0.3100 | 1.0000 | 2.7103 | 0.8253 | 10.8539 |
| Aug | St. Jerome | MGE (mixture estimation) | 0.2800 | 0.8190 | 1.8998 | 1.0000 | 10.3986 |
| Sep | Dorval | MGE (mixture estimation) | 0.7900 | 0.8116 | 7.7097 | 1.0000 | 19.3866 |
| Sep | Cornwall | MGE (mixture estimation) | 0.9400 | 0.7182 | 10.8961 | 1.0000 | 31.8589 |
| Sep | Drummondville | MEW (mixture estimation) | 0.0800 | 1.0000 | 1.0063 | 0.8157 | 8.6807 |
| Sep | Farnham | MGE (mixture estimation) | 0.9300 | 0.8264 | 10.8825 | 1.0000 | 0.0475 |
| Sep | Lennoxville | MGE (mixture estimation) | 0.9300 | 0.8900 | 8.8234 | 1.0000 | 0.0432 |
| Sep | Morrisburg | MEW (mixture estimation) | 0.2900 | 1.0000 | 4.5615 | 0.7487 | 9.3987 |
| Sep | Oka | Mixed Exponential (ML estimation) | 0.1868 | 1.0000 | 0.8454 | 1.0000 | 9.9319 |
| Sep | Ottawa CDA | MEW (mixture estimation) | 0.0800 | 1.0000 | 0.0693 | 0.8565 | 6.8425 |
| Sep | St. Alban | Mixed Exponential (mixture estimation) | 0.3400 | 1.0000 | 2.5006 | 1.0000 | 13.5342 |
| Sep | St. Jerome | MEW (mixture estimation) | 0.0500 | 1.0000 | 0.1065 | 0.8816 | 8.5960 |
| Oct | Dorval | Mixed Exponential (mixture estimation) | 0.0500 | 1.0000 | 0.0378 | 1.0000 | 6.7779 |
| Oct | Cornwall | MGE (mixture estimation) | 0.5000 | 0.7628 | 11.3358 | 1.0000 | 4.7850 |
| Oct | Drummondville | MEW (mixture estimation) | 0.0500 | 1.0000 | 0.0563 | 0.9837 | 7.3368 |
| Oct | Farnham | MGE (mixture estimation) | 0.9800 | 0.8817 | 7.1917 | 1.0000 | 28.2774 |
| Oct | Lennoxville | Mixed Exponential (mixture estimation) | 0.8100 | 1.0000 | 8.2191 | 1.0000 | 2.4109 |
| Oct | Morrisburg | MEW (mixture estimation) | 0.0600 | 1.0000 | 0.0967 | 0.8458 | 6.4358 |
| Oct | Oka | MEW (mixture estimation) | 0.6600 | 1.0000 | 5.9555 | 0.6974 | 6.0077 |
| Oct | Ottawa CDA | MEW (mixture estimation) | 0.0300 | 1.0000 | 0.0455 | 0.8816 | 6.0459 |
| Oct | St. Alban | MEW (mixture estimation) | 0.0700 | 1.0000 | 0.5527 | 0.9711 | 8.3134 |
| Oct | St. Jerome | MEW (mixture estimation) | 0.8200 | 1.0000 | 7.0773 | 0.6679 | 7.2162 |
| Nov | Dorval | MEW (mixture estimation) | 0.0300 | 1.0000 | 0.0402 | 0.9967 | 6.8390 |
| Nov | Cornwall | MGE (mixture estimation) | 0.9000 | 0.9070 | 7.6800 | 1.0000 | 3.6425 |
| Nov | Drummondville | Mixed Exponential (ML estimation) | 0.0570 | 1.0000 | 0.0519 | 1.0000 | 7.6304 |
| Nov | Farnham | MGW (ML estimation) | 0.4847 | 0.6513 | 5.3140 | 1.3761 | 9.5088 |
| Nov | Lennoxville | MEW (mixture estimation) | 0.0200 | 1.0000 | 0.0849 | 0.8793 | 5.5683 |
| Nov | Morrisburg | MEW (mixture estimation) | 0.0400 | 1.0000 | 0.0670 | 0.9560 | 7.0175 |
| Nov | Oka | MEW (mixture estimation) | 0.0200 | 1.0000 | 0.0542 | 0.9742 | 7.3080 |
| Nov | Ottawa CDA | MEW (mixture estimation) | 0.0700 | 1.0000 | 0.0604 | 0.9530 | 6.3790 |
| Nov | St. Alban | Exponential (ML estimation) | | 1.0000 | 7.3721 | 1.0000 | 7.3721 |
| Nov | St. Jerome | Mixed Exponential (mixture estimation) | 0.0600 | 1.0000 | 0.4588 | 1.0000 | 7.6587 |
| Dec | Dorval | MGW (mixture estimation) | 0.8000 | 0.8985 | 7.7548 | 0.8522 | 1.0771 |
| Dec | Cornwall | Gamma (ML estimation) | | 0.8046 | 7.3382 | | |
| Dec | Drummondville | MEW (mixture estimation) | 0.0200 | 1.0000 | 0.0554 | 0.8840 | 5.6135 |
| Dec | Farnham | MEW (mixture estimation) | 0.0700 | 1.0000 | 0.0502 | 0.9194 | 5.7446 |
| Dec | Lennoxville | MEW (mixture estimation) | 0.0500 | 1.0000 | 0.0412 | 0.8522 | 5.5864 |
| Dec | Morrisburg | MEW (mixture estimation) | 0.0500 | 1.0000 | 0.0487 | 0.9386 | 5.9615 |
| Dec | Oka | Mixed Exponential (mixture estimation) | 0.0900 | 1.0000 | 1.6349 | 1.0000 | 6.6476 |
| Dec | Ottawa CDA | Mixed Exponential (mixture estimation) | 0.1400 | 1.0000 | 0.5695 | 1.0000 | 5.7101 |
| Dec | St. Alban | Mixed Exponential (mixture estimation) | 0.3100 | 1.0000 | 2.1700 | 1.0000 | 7.6884 |
| Dec | St. Jerome | MEW (mixture estimation) | 0.0300 | 1.0000 | 0.4504 | 0.9038 | 7.0180 |

Table 3. The model selection result for the modeling of the daily precipitation amounts.

The final model of Table 3 simply selects the candidate distribution that has the maximum p-value of the likelihood ratio test among other candidate distributions. When the p-values of all the candidate distributions are smaller than 0.05, the ML estimation of the MGW distribution is selected. When candidate distributions have larger ML values than the corresponding ML estimation of the MGW distribution, the AIC is adopted to select the final model only from these candidate distributions without the consideration of the p-values of the likelihood ratio tests of other candidate distributions. When the ML estimation of the MGW distribution results in the ML estimation of the Gamma/Weibull distribution, the AIC is applied to select the final model from these three models: the Gamma/Weibull distribution with the ML estimation, the mixed exponential distribution with the ML estimation, and the MGW distribution with the mixture estimation.

As shown in Table 2, the likelihood ratio test under the MGW distribution framework gives a result that is consistent with previous research that the appropriate model for daily precipitation amounts falls between the Gamma and Weibull distributions with the ML estimation approach and favors the exponential distribution with the ML estimation approach in certain situations. As also shown is Table 2, the mixed exponential distribution with the ML estimation approach, which serves as a model that fills the gap between the Gamma and Weibull distributions with the ML estimation approach, is an appropriate model for most of the cases. The appropriateness of this model is also consistent with previous research. Finally, Table 3 shows that there is a necessity to have a probability distribution that is a slightly more flexible than the mixed exponential distribution to fill the deficit of the mixed exponential distribution in the modeling of daily precipitation amounts. The MGW distribution serves as an appropriate solution.

**Conclusions**

In this study, the MGW distribution was introduced for the daily precipitation amount modeling, and its application to the precipitation record of southern Quebec was shown as an example. The following conclusions can be drawn:

1. The parameter setting of the MGW distribution was shown to be able to integrate the most commonly used probability distributions for the daily precipitation amount modeling, i.e. the exponential, Gamma, Weibull, and mixed exponential distributions, into one framework. The exponential, Gamma, Weibull, and mixed exponential distributions can all be viewed as special cases of the MGW distribution, and, therefore, the MGW distribution is an ideal probability distribution for further consideration in the selection of a suitable distribution of daily precipitation amounts.

2. The mixture estimation approach was introduced for estimating the parameters of the MGW distribution in modeling daily precipitation amounts. The MGW distribution with the mixture estimation was able to provide a more flexible fit of the tail probability than the commonly used probability distributions while forcing the mean and variance of the model equal to the mean and variance of the observed daily precipitation amounts. Thus, the MGW distribution with the mixture estimation could be considered as a flexible and general model for representing the distribution of daily precipitation amounts.

3. The ML estimation approach for the MGW distribution was developed. By also taking the physical properties of daily precipitation amount records into consideration, the developed ML estimation approach is able to give an estimation result that is both statistically and physically reasonable to the modeling of daily precipitation amounts. Finally, the ML estimation of the MGW distribution serves an important role in the model selection of the daily precipitation amount modeling.

4. Based on the large sample properties of likelihood ratios and the parameter setting of the MGW distribution, a model selection criterion for daily precipitation amount modeling was introduced to integrate all the candidate models into one general framework. The application of the model selection criterion gives a reasonable comparison result that is consistent with previous research, and it further implies that the MGW distribution is an appropriate solution for filling the limitation of the mixed exponential distribution in the modeling of daily precipitation amounts.

In summary, the MGW distribution serves not only as a new probability distribution for the daily precipitation amount modeling but also as a model selection framework that incorporates the commonly used probability distributions for the modeling of daily precipitation amounts into one general and consistent framework.

**Acknowledgements**
I appreciate the financial support of McGill Engineering Doctoral Awards (MEDA) provided by the Faculty of Engineering of McGill University. Also, I’m grateful for the funding of the Natural Sciences and Engineering Research Council of Canada (NSERC) via the Research Assistantship awarded. The support of each party was integral during the production of this manuscript.